\documentclass[12pt]{article}

\usepackage{epsfig}

\begin{document}

\title{\ Two diffraction cones of elastic scattering and structural symmetry conjecture}

\author{A.P.~Samokhin  \\
\textit{A.A. Logunov Institute for High Energy Physics}\\
\textit{of NRC ``Kurchatov Institute''}\\
\textit{Protvino, 142281, Russian Federation}}

\date{}

\maketitle

\begin{abstract}
The energy dependence of the differential cross-section of elastic proton-proton scattering in the ISR--LHC energy range is discussed for fixed values 
of momentum transfer in the region of the forward diffraction cone, in the region beyond the second maximum of $d\sigma/dt$ and in the vicinity 
of the dip-bump structure. As can be seen from the currently available experimental data, the differential cross-section in the region beyond 
the second maximum has exactly the same properties as in the forward diffraction cone, including the existence of a stationary point, that is, 
we observe a second diffraction cone, which has the same origin as the forward peak. The simplest natural explanation for this experimental fact is that the amplitude of high energy elastic scattering
is the sum of two similar terms that have the same status and differ only in the values of parameters. The energy dependence of $d\sigma/dt$ for a 
fixed $t$ in the region of the dip-bump structure, where these two terms interfere, confirms the above observations. We discuss the possible origin 
of the two-component structure of the high energy elastic scattering amplitude and explain this by the structural symmetry of the amplitude.
\end{abstract}

\textit{Keywords:} Elastic scattering; Differential cross-section; Diffraction cones; Stationary points; Structural symmetry

\section{Introduction}

The properties of high energy hadron-hadron elastic scattering are determined not only by the complex
nature of the interaction of extended hadrons, but also, due to the unitarity, by the totality
of all open inelastic channels. For this reason, elastic scattering has unique, often unexpected properties.
This article discusses one of these unusual properties of an elastic differential cross-section.

At a fixed energy in the ISR--LHC energy range, the differential cross-section $d\sigma(s,t)/dt$ of elastic $pp$ and $p\bar{p}$ scattering
has the following main features [1-22]. For small momentum transfer values $t$, a forward diffraction cone (or forward peak) is observed,
that is, a fast, rather structureless, approximately exponential fall-off of $d\sigma/dt$ with $|t|$. Then follows the region of the dip-bump
structure (the shoulder for $p\bar{p}$), where $d\sigma/dt$ has a local minimum (dip) and a second maximum (bump). In the region beyond the 
second maximum, a structureless exponential drop of $d\sigma/dt$ is also observed, but with a much smaller slope than at the forward peak.

This structure of the differential cross-section is energy dependent. The slope $B(s)$ of the forward diffraction cone and its height
$(d\sigma/dt)_{t=0}$ grow with energy, i.e. there is a shrinkage of the forward peak. As the energy increases, the dip-bump structure moves 
to $t=0$, its width decreases, and the values of $d\sigma/dt$ at its points grow. In the region beyond the second maximum, 
the differential cross-section is practically independent of energy when $23.5<\sqrt{s}<62.5$ GeV (ISR energy range) [3], but the energy dependence
of $d\sigma/dt$ in this region of $t$ becomes apparent when comparing ISR data with TOTEM 7 and 13 TeV data [4,7].

As is known, the existence of a forward peak in the differential cross-section follows from the general principles of local quantum field theory 
(QFT), such as unitarity and analyticity (i.e., causality and the short-range nature of strong interaction) [23-27]. This means that the hadron-hadron
elastic scattering near the forward direction looks as if hadrons were point-like structureless objects.

The nature of the behaviour of $d\sigma/dt$ in the region beyond the second maximum is still unknown. In models based on the Regge theory, 
this is associated with double or multiple Pomeron exchange [28]. In such models, as a rule, it is difficult to avoid the appearance of secondary 
(experimentally unobservable) structures. For a long time, the three-gluon exchange model was popular [29,30], which predicts the energy-independent
$|t|^{-8}$ behaviour of $d\sigma/dt$ at large momentum transfer. The currently observed energy dependence of $d\sigma/dt$ proves the inapplicability 
of this model in the region under discussion beyond the second maximum. In general, the values of $t$ available at the LHC are too small to observe
the behaviour due to hard scattering among the proton constituents, in particular, to observe the dimensional counting regime [31,32]. 
There are models in which the behaviour beyond the second maximum of $d\sigma/dt$ is associated with the assumed layered (having several scales) 
structure of the proton or its interaction region [33-39]. The fact that the local slope of 
$d\sigma/dt$ in this region is about 4 times smaller than in the forward diffraction cone is interpreted as the presence of a new internal structure inside the proton [36-38] (or as the presence of some substructure [35]), which is half the size of the proton. If this is so, then the transition from the external structure to the internal one should be smooth, as is the case for Rutherford scattering by atoms or for the $p_{T}$-dependence of the inclusive spectrum of hadrons (the transition from hadron to its quark-gluon structure). But instead of a smooth $t$-dependence of the differential cross-section, a dip-bump structure is observed. On the other hand, it is known that the differential cross-section for elastic nucleus-nucleus scattering at small momentum transfer is very sensitive to the internal structure of colliding nuclei. According to Glauber-type calculations for elastic scattering of composite systems, a single diffractive minimum occurs only for two-component colliding objects, as in elastic deuteron-deuteron scattering, and a much slower decrease of $d\sigma/dt$ in the region beyond the diffractive minimum is due to multiparticle collisions of constituents [40]. The quark-diquark model assumes that the nucleon is composed of two constituents, a quark and a diquark, and that elastic nucleon-nucleon scattering can be regarded exactly as elastic deuteron-deuteron scattering in nuclear physics [41]. A comparison of the predictions of this model with the ISR data for the $pp$ differential cross-section in $0<|t|<3$ GeV$^{2}$ range allows one to determine the model parameters -- the sizes of the quark, diquark and the distance between them in the proton [41]. The real extended and unitarized version of this quark-diquark model describes the ISR and 7 TeV TOTEM data [42], as well as the TOTEM data at $\sqrt{s} = 2.76$ TeV and the $p\bar{p}$ differential cross-section data at $\sqrt{s} = 0.546$ and 1.96 TeV, but cannot describe the 13 TeV TOTEM data [43]. Note that attempts to take into account the quark-quark structure of the diquark give additional diffractive minima and destroy the ability of this model to describe experimental data [42].

To describe the dip-bump structure of $d\sigma/dt$, it was proposed to parameterize the amplitude of high energy elastic scattering as the sum 
of two exponentials with a relative phase [44]. This gives a simple and natural description of the ISR data over a wide $t$ range ($0.6<|t|<5.1$
GeV$^{2}$) [2,3], as well as LHC 7 TeV data (in $0.38<|t|<2.4$ GeV$^{2}$) [45-47], including a part of the forward diffraction cone, the dip-bump 
structure, and the region beyond the second maximum. To describe the data in the region of small $t$, the proton form factor was introduced 
as a multiplier in front of the first exponential [46]. This modified Phillips and Barger parameterization describes the behaviour of $d\sigma/dt$ 
at ISR and LHC energies over the entire $t$ range [48].

In models with a two-component structure of the high energy elastic scattering amplitude, there are different interpretations of the first and
second terms: the Pomeron and a non-Regge background [44], the dipole Pomeron and dipole Odderon [49], the soft Pomeron and hard Pomeron [50],
Regge pole theory parameterization of both terms [47], the Pomeron pole plus the gray disk model [51]. We see that the nature of the two-component
amplitude of high energy elastic scattering is an open question.

The shrinkage of the forward diffraction cone for $pp$ elastic scattering has a striking feature: the value of $d\sigma/dt$ at $t=t_{\ast}\approx - 0.21$ GeV$^{2}$ is practically independent of the energy in the ISR--LHC energy range. This \textit{stationary point} of the differential cross-section was observed [52] when only ISR data [1-3] and 7 TeV TOTEM data for $|t|\leq2.4$ GeV$^{2}$ [4,5] were available. Shortly thereafter, the TOTEM Collaboration presented preliminary \textit{unnormalized} 13 TeV data for $d\sigma/dt$ in the range of $0.05<|t|<3.4$ GeV$^{2}$ [53]. We normalized this data at the point $t=t_{\ast}$ and thus 
obtained a prediction for $d\sigma/dt$ at $\sqrt{s} = 13$ TeV over the entire range of $0.05<|t|<3.4$ GeV$^{2}$ [54]. This allowed us to notice that the energy behaviour of $d\sigma/dt$ in the region beyond the second maximum is very similar to its behaviour in the forward diffraction cone, in particular, $d\sigma/dt$ reveals a \textit{stationary point} at $t=t_{\ast \ast}\approx - 2.3$ GeV$^{2}$ in the region beyond the second maximum, just like it has a stationary point at $t=t_{\ast}$ in the forward diffraction cone. In other words, in the region beyond the second maximum, we observe the \textit{second diffraction cone}, which has the same origin as the forward diffraction cone. The existence of two similar diffraction cones of $d\sigma/dt$ motivates the following natural assumption: the high energy elastic scattering amplitude is the sum of two similar terms that have the same status, the same analytic structure, and differ only in the values of the parameters. The interference of these terms gives a dip-bump structure of the differential cross-section [54].

In this paper, we analyze the energy dependence of $d\sigma/dt$ in the ISR--LHC energy range for fixed values of the momentum transfer using all currently available experimental data for $pp$ elastic scattering [1-13], including recent TOTEM data at $\sqrt{s} = 13$ and 2.76 TeV [6,7;8], data from STAR Collaboration at $\sqrt{s} = 200$ GeV [13], as well as data for $p\bar{p}$ from SPS [17-19] and Tevatron [20-22]. A detailed and complete explanation of the phenomenon of approximate stationarity of $d\sigma/dt$ as a consequence of the correlated growth of the total cross-section $\sigma_{\mathrm{tot}}(s)$ and the slope $B(s)$ is given. All predictions of Ref. [54] are confirmed, that is, in the range $0<|t|<4$ GeV$^{2}$ there are indeed two shrinking diffraction cones of $d\sigma/dt$, which have approximately stationary points at $t=t_{\ast}$ and $t=t_{\ast \ast}$ in the ISR--LHC energy range. An analysis of the energy dependence of $d\sigma/dt$ for a fixed $t$ in the region of the dip-bump structure is carried out, which confirms that in this region there is a transition from the first shrinking diffraction cone to the second.

The possible origin of a non-trivial algebraic structure of the elastic scattering amplitude is also discussed. Each of the three physical 
channels of this elastic process is characterized by a certain set of parameters. On the other hand, these channels have the same status,
since the amplitude describes each of them on an equal footing. Therefore, it is natural to assume that the elastic scattering amplitude is
a symmetric function for permutations of kinematic variables along with their sets of channel parameters (recall that the permutation 
of only kinematic variables gives the amplitude in the crossed or in the exchange channel). If the elastic scattering amplitude has such a 
symmetry, hereinafter referred to as \textit{structural symmetry}, then in the general case it should be a composition of irreducible 
representations of the permutation group $S_{3}$. All six terms of this composition (the TREE-amplitude) have the same status and are equally 
important at low energies, but only two of them dominate at high energies in the diffraction region. The latter can be seen from simple explicit
examples, but in the general case this is a conjecture. We anticipate that this property of the TREE-amplitude can be proved as a consequence of 
analyticity, unitarity, and other constraints of local QFT. Thus, in this context, the existence of two diffraction cones of $d\sigma/dt$ 
can be considered as a manifestation of the structural symmetry of the elastic scattering amplitude.

In Section 2, we analyze the energy dependence of $d\sigma/dt$ for fixed values of the momentum transfer in the first diffraction cone. In Section 3,
we discuss the behaviour of $d\sigma/dt$ in the second diffraction cone, the two-component structure of the high energy elastic scattering amplitude, 
and the energy dependence of $d\sigma/dt$ for a fixed $t$ in the dip-bump region. The structural symmetry of the elastic scattering amplitude is 
discussed in Section 4. A brief summary and discussion are given in Section 5.

\section{Energy dependence of $d\sigma/dt$ at fixed values of momentum transfer}

Looking at Fig. 1, it is easy to see that the differential cross-section for $pp$ elastic scattering has a striking feature:
its values at $t\approx - 0.21$ GeV$^{2}$ and $t\approx - 2.3$ GeV$^{2}$ are practically energy-independent in the ISR--LHC energy range. In other
words, the differential cross-section $ d\sigma(s,t)/dt$ has \textit{stationary points} at $t_{\ast}\approx - 0.21$ GeV$^{2}$ and
$t_{\ast \ast}\approx - 2.3$ GeV$^{2}$ [54].
\begin{figure}[t]
\centering
\includegraphics[height=8.5cm]{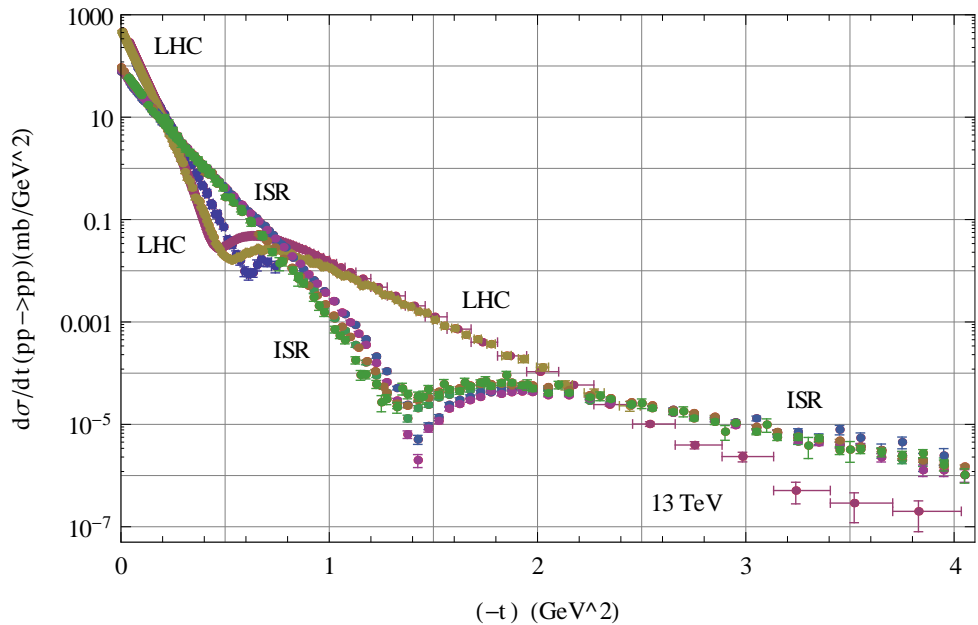}
\caption{Differential cross-section for $pp$ elastic scattering at the ISR ($ \sqrt{s} = 23.5, 30.7, 44.7, 52.8, 62.5$ GeV) and LHC 
($ \sqrt{s} = 2.76, 7, 13$ TeV) energies. Data are taken from Refs. [1-8].}
\end{figure}
\begin{figure}[t]
\centering
\includegraphics[height=8.5cm]{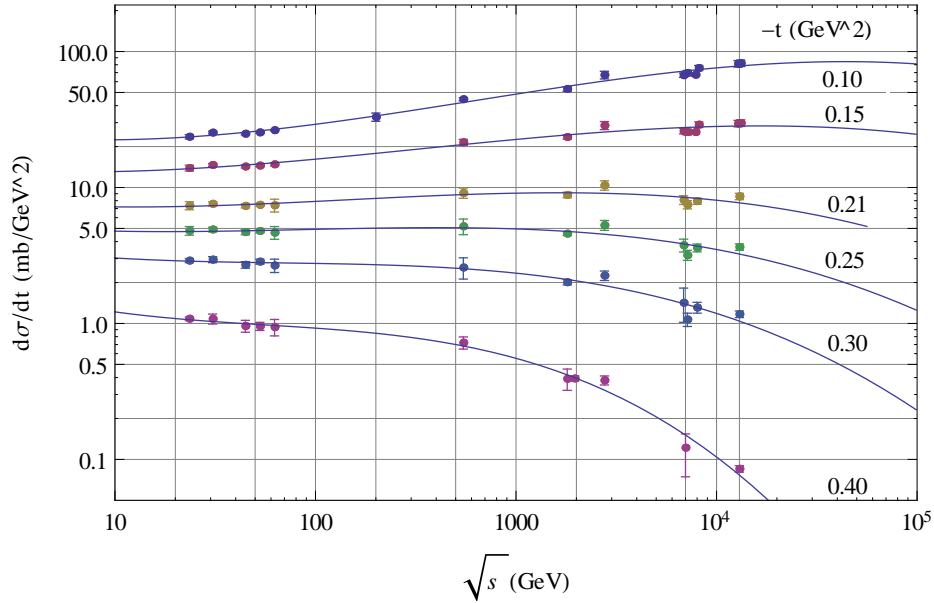}
\caption{Energy dependence of the elastic differential cross-section at fixed values of the momentum transfer $t$ in the first diffraction cone. 
Data are taken from Refs. [1-13] and [17-22]. The lines are the polynomial fits to the data points.}    
\end{figure}

A more detailed picture of this phenomenon is seen from the energy dependence of $ d\sigma(s,t)/dt$ for fixed values of momentum transfer
in the vicinity of the first stationary point $t_{\ast}$ (see Fig. 2). Since the high energy differential cross-sections for $pp$ and 
$p\bar{p}$ elastic scattering noticeably differ only in the dip-bump region [15,16], [8,55], we took into account, along with the 
$pp$ data [1-13], data for $p\bar{p}$ from SPS [17-19] and Tevatron [20-22]. From Fig. 2, we can see that the shrinkage of the forward
diffraction cone in the ISR--LHC energy range occurs in some specific way. For $t=-0.40$ GeV$^{2}$, the differential cross-section decreases with 
energy, at $t=-0.30$ GeV$^{2}$ it remains practically unchanged up to $\approx500$ GeV, and then decreases with increasing energy. 
At $t=-0.25$ GeV$^{2}$, $ d\sigma(s,t)/dt$ remains almost unchanged up to $\approx3$ TeV, and then decreases. For $t=-0.21$ GeV$^{2}$, 
the differential cross-section has a local minimum at $ \sqrt{s}\approx15$ GeV, then slowly grows up to $\approx2$ TeV, where it has a local maximum
and then slowly decreases, but the value of $ d\sigma(s,t)/dt$ remains almost constant throughout the ISR--LHC energy range. 
At $t=-0.15$ GeV$^{2}$, the differential cross-section noticeably increases from ISR to LHC, where it reaches a local maximum.
At last, $ d\sigma(s,t)/dt$ grows rapidly with energy in the ISR--LHC energy range at $t=-0.10$ GeV$^{2}$.

What is the reason for such nontrivial energy behaviour of $ d\sigma(s,t)/dt$ at fixed values of momentum transfer? According to the optical
theorem (which is a consequence of the unitarity condition), the exact expression for the differential cross-section is given by
\begin{equation}
   \frac{d\sigma(s,t)}{dt} = \frac{\sigma_{\mathrm{tot}}^{2}(s)(1+\rho^{2}(s))}{16\pi}\exp[\int\limits_{0}^{t}dt^{'}B(s,t^{'})],  
\end{equation}
where $\sigma_{\mathrm{tot}}(s)$ is the total cross-section, $\rho(s) = ReT(s,0)/ImT(s,0)$ is the ratio of the real to the imaginary part of the 
forward amplitude of elastic nuclear scattering, and $B(s,t)= (d/dt)\ln(d\sigma(s,t)/dt)$ is the local slope parameter. 
Since $ \rho^{2}(s)$ is small and in the forward peak region ($0<-t<0.4$ GeV$^{2}$) the local slope $B(s,t)\approx B(s,t=0)\equiv B(s)$,
we have an approximate formula
\begin{equation}
   \frac{d\sigma(s,t)}{dt} \approx \frac{\sigma_{\mathrm{tot}}^{2}(s)}{16\pi}\exp[tB(s)].
\end{equation}
Then for a fixed value of $t=t_{1}$ we have
\begin{equation}
 (\ln\frac{d\sigma}{dt})_{t=t_{1}}^{\prime} = (2\frac{\sigma_{\mathrm{tot}}^{\prime}}{\sigma_{\mathrm{tot}}} - |t_{1}|B^{\prime}),
\end{equation}
where the prime denotes the derivative with respect to $ \ln (\sqrt{s})$. The forward slope for elastic scattering $pp$, $B(s)$, is an increasing
function of energy, that is, $B^{\prime}>0$ for all energies. From ISR to TeV energy $B^{\prime}\approx const $, but then it grows with
energy [14]. Let us remind that the high energy growth of $B(s)$ is a unitary-motivated consequence of the growth of the total cross-section [56].

At low energies (up to ISR) $\sigma_{\mathrm{tot}}(s)$ decreases with energy, therefore $\sigma_{\mathrm{tot}}^{\prime}<0$ and
in accordance with formula (3), $ d\sigma(s,t)/dt$ will decrease with energy for any fixed $t_{1}$. Indeed, we know [57] that this is so. Starting 
from ISR, where $\sigma_{\mathrm{tot}}(s)$ begins to grow, the first term in the brackets of Eq. (3) becomes positive, and the sign of the 
derivative, $(d\sigma/dt)_{t=t_{1}}^{\prime}$, will depend on the value of the fixed $|t_{1}|$. For small values of $|t_{1}|$ the first positive term
dominates and the differential cross-section grows, for large values of $|t_{1}|$ the second term dominates and $ d\sigma/dt$ decreases with energy.
For some intermediate values of fixed $|t_{1}|$, these terms approximately compensate each other and $ d\sigma/dt\approx const$ in a certain
energy range.
\begin{figure}[t]
\centering
\includegraphics[height=8.5cm]{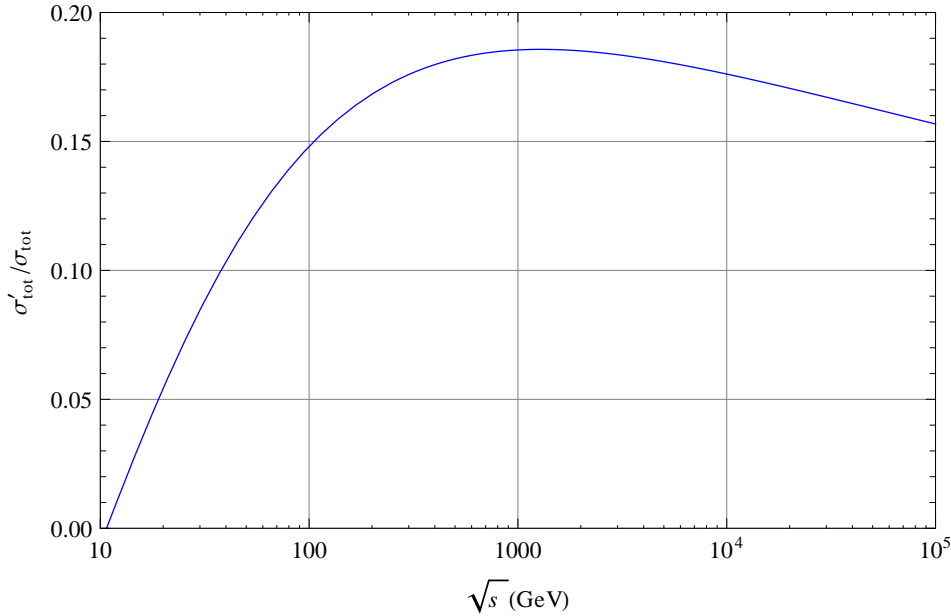}
\caption{Energy dependence of the ratio $\sigma_{\mathrm{tot}}^{\prime}/\sigma_{\mathrm{tot}}$. The COMPETE Collaboration fit [58,59] is used
for a function $\sigma_{\mathrm{tot}}(s)$.}    
\end{figure}
\begin{figure}[t]
\centering
\includegraphics[height=8.5cm]{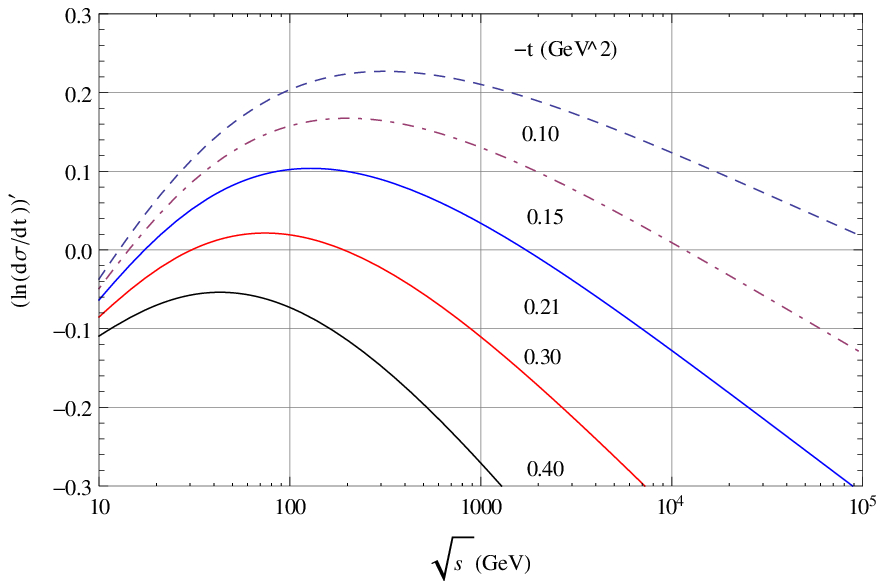}
\caption{Energy dependence of the difference $(2\sigma_{\mathrm{tot}}^{\prime}/\sigma_{\mathrm{tot}} - |t_{1}|B^{\prime})$ at fixed values
of $|t_{1}|$. The fits [58,59] and [60] are used for the total cross-section $\sigma_{\mathrm{tot}}(s)$ and forward slope $B(s)$ respectively.}
\end{figure}

To see the fine structure of this behaviour, we can use the fits for $\sigma_{\mathrm{tot}}(s)$ [58,59] and slope $B(s)$ [60,61], which describe the 
data in the ISR--LHC energy range. The ratio $\sigma_{\mathrm{tot}}^{\prime}/\sigma_{\mathrm{tot}}$ first increases, reaches a maximum at TeV energy,
and then decreases (see Fig. 3). By the way, we note that the slope $B(s)$, the ratios $\sigma_{\mathrm{el}}/\sigma_{\mathrm{tot}}$,
$(\sigma_{\mathrm{inel}}-\sigma_{\mathrm{el}})/B $ also change their behaviour in the TeV energy range [14,62]. The curves in Fig. 4 give the energy 
dependence of the difference $(2\sigma_{\mathrm{tot}}^{\prime}/\sigma_{\mathrm{tot}} - |t_{1}|B^{\prime})$ for fixed values of $|t_{1}|$. We see that
for $t_{1}=-0.40$ GeV$^{2}$ this difference is negative and, therefore, according to Eq. (3) the differential cross-section decreases in the ISR--LHC 
energy range. When $t_{1}=-0.30$ GeV$^{2}$, the differential cross-section has a local minimum at $ \sqrt{s}\approx30$ GeV, then it increases very 
slowly, reaches a local maximum at $ \sqrt{s}\approx200$ GeV, and then decreases, but the value of the derivative is close to zero up to 
$\approx500$ GeV. For $t_{1}=-0.21$ GeV$^{2}$ the local minimum and maximum lie at $\approx15$ GeV and $\approx2$ TeV respectively, the value 
of the derivative is small and $d\sigma/dt\approx const$ from ISR energies up to 20 TeV. For $t_{1}=-0.15$ GeV$^{2}$ the local minimum and maximum 
lie at $\approx13$ GeV and $\approx10$ TeV respectively, the difference $(2\sigma_{\mathrm{tot}}^{\prime}/\sigma_{\mathrm{tot}} - |t_{1}|B^{\prime})$ 
has a noticeable value and $d\sigma/dt$ significantly increases from ISR energies up to LHC. The value of the derivative for $t_{1}=-0.10$ GeV$^{2}$ 
is large, and the differential cross-section rapidly grows with energy in the ISR--LHC energy range. 

So, we see that the energy dependence of $(2\sigma_{\mathrm{tot}}^{\prime}/\sigma_{\mathrm{tot}} - |t_{1}|B^{\prime})$ for fixed values of $|t_{1}|$,
see Fig. 4, explains all the details of the $d\sigma/dt$ behaviour in a forward diffraction cone, including the approximate stationarity (see Fig. 2). 
In other words, this behaviour is a consequence of the correlated growth of $\sigma_{\mathrm{tot}}(s)$ and $B(s)$.

Of course, the stationarity and growth of the differential cross-section at a fixed $0<|t_{1}|<0.21$ GeV$^{2}$ are 
the ``transitory'' properties of $d\sigma/dt$. Due to the growth of the total cross-section, the ratio 
$\sigma_{\mathrm{tot}}^{\prime}/\sigma_{\mathrm{tot}}$ will decrease to zero at high energies, hence the difference 
$(2\sigma_{\mathrm{tot}}^{\prime}/\sigma_{\mathrm{tot}} - |t_{1}|B^{\prime})$ will be negative at sufficiently high energies for any fixed value of 
$|t_{1}|$ and, therefore, according to Eq. (3), $(d\sigma/dt)_{t=t_{1}}$ will decrease with energy. For example, $d\sigma/dt$ at fixed 
$t_{1}=-0.21$ GeV$^{2}$ will noticeably decrease starting from 20-30 TeV (see Fig. 2 and [54]). The differential cross-section at fixed 
$t_{1}=-0.15$ GeV$^{2}$, which practically does not change from 3 TeV to 40 TeV, will noticeably decrease starting from 50-60 TeV, etc. 
This means that the value of $|t_{1}|$, at which in a certain energy range $(d\sigma/dt)_{t=t_{1}}\approx const$, decreases with increasing energy.

Thus, $(d\sigma/dt)_{t=t_{1}}$ decreases (or will decrease) with energy at any fixed value of $|t_{1}|>0$ from the forward diffraction cone. 
But this decrease is also transitory, because there is a second diffraction cone.
\begin{figure}[t]
\centering
\includegraphics[height=8.5cm]{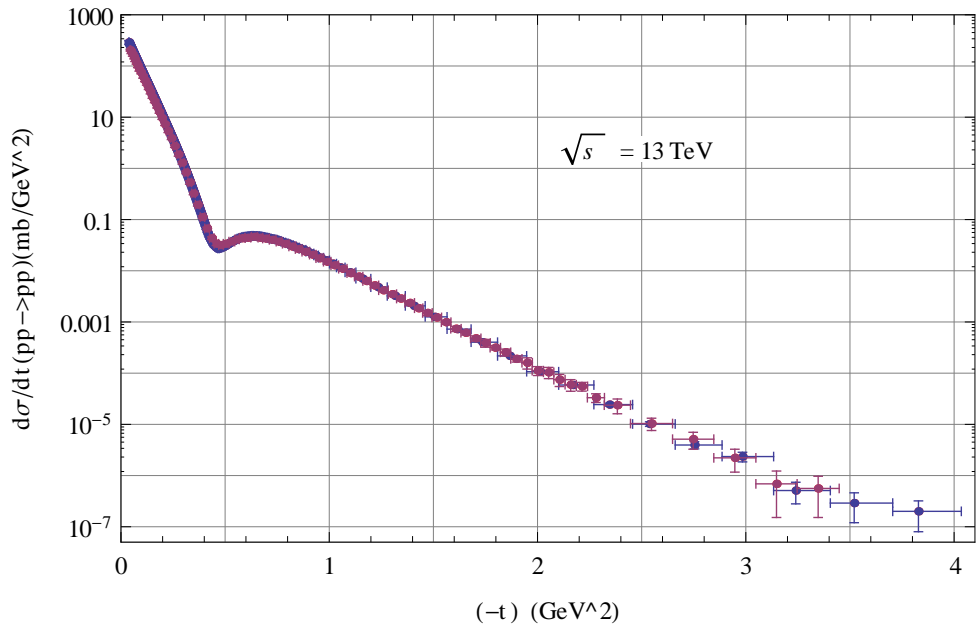}
\caption{
Comparison of the preliminary unnormalized 13 TeV data for $d\sigma/dt$ [53] in our normalization [54] with the currently published normalized 
TOTEM data for $d\sigma/dt$ at $\sqrt{s} = 13$ TeV [6,7].}
\end{figure}

\section{Second diffraction cone and structure of high energy elastic scattering amplitude}

The existence of a stationary point of the differential cross-section for $pp$ elastic scattering at the point $t_{\ast}\approx - 0.21$ GeV$^{2}$
was observed [52] when only ISR data [1-3] and 7 TeV data for $|t|\leq2.4$ GeV$^{2}$ [4,5] were available. Soon thereafter, 
the TOTEM Collaboration showed the preliminary unnormalized 13 TeV data [53] for $d\sigma/dt$ in the range of $0.05<|t|<3.4$ GeV$^{2}$. 
These data were normalized at the point $t=t_{\ast}$, and thus, we obtained the prediction for $d\sigma/dt$ at 13 TeV in the whole range of
$0.05<|t|<3.4$ GeV$^{2}$ [54]. A comparison of this prediction with the currently published normalized TOTEM data [6,7] is shown in Fig. 5. 
As can be clearly seen, there is an obvious agreement between them. So, all currently available experimental data, including 2.76 TeV data [8],
confirm that $d\sigma/dt$ at $t=t_{\ast}$ practically does not change in the ISR--LHC energy range, and confirm the above
self-consistent picture of shrinkage of the forward diffraction cone (see Fig. 1 and Fig. 2). In addition, preliminary 13 TeV data, normalized 
at the point $t=t_{\ast}$, allowed us to predict the existence of a second stationary point of $d\sigma/dt$ at $t=-2.3$ GeV$^{2}$ [54].
\begin{figure}[t]
\centering
\includegraphics[height=8.5cm]{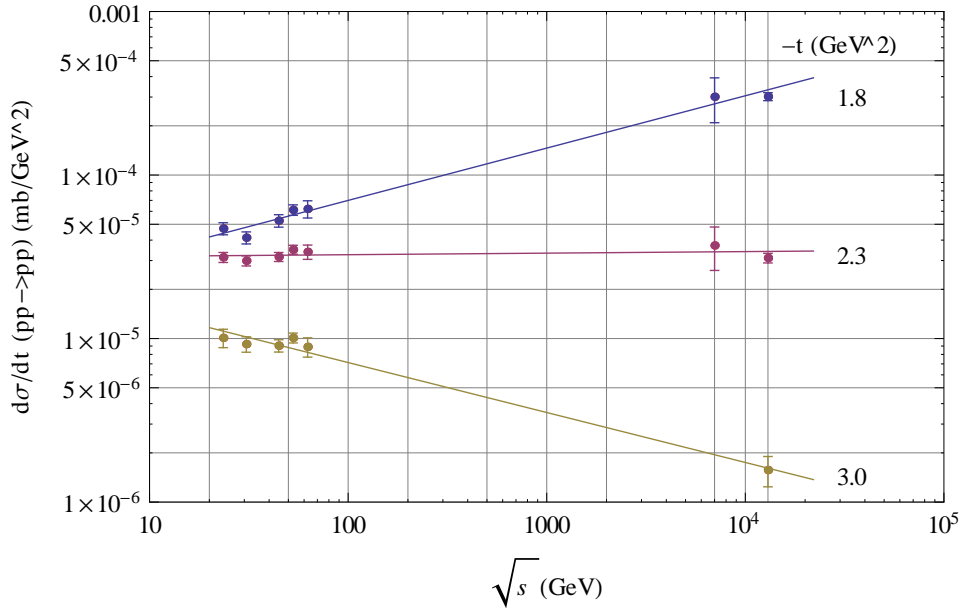}
\caption{Energy dependence of the differential cross-section for $pp$ elastic scattering at fixed values of the momentum transfer $t$ 
in the second diffraction cone. Data are taken from Refs. [1-7]. The lines are the linear least-square fits to the data points.}
\end{figure}

At low energies (up to ISR) and fixed values of the momentum transfer from the interval $2\leq|t|\leq5$ GeV$^{2}$, the differential
cross-section for $pp$ elastic scattering decreases with energy [57]. In the ISR energy range ($23.5<\sqrt{s}<62.5$ GeV), $d\sigma/dt$ for
$2.3\leq|t|\leq5.1$ GeV$^{2}$ is well approximated by an exponential form $C\exp(-D|t|)$, where the parameters $C$ and $D$ are practically energy
independent [3] (see Fig. 1). This approximate energy independence of $(d\sigma/dt)_{t=t_{2}}$ at a fixed $t=t_{2}$ in the region beyond the second 
maximum resembles the behaviour of $(d\sigma/dt)_{t=t_{1}}$ at a fixed $t=t_{1}$ in the forward diffraction cone (compare Fig. 2 and Fig. 6 in the 
ISR energy range). Comparing the ISR data with the TOTEM 7 and 13 TeV data at a fixed $t=t_{2}$ in the vicinity $|t|=2.5$ GeV$^{2}$ (see Fig. 6), 
we can observe exactly the same trend as in the forward diffraction cone (see Fig. 2): for fixed values of $|t|<2.3$ GeV$^{2}$ the differential 
cross-section grows with energy, while for $|t|>2.3$ GeV$^{2}$ it decreases. At $t=-2.3$ GeV$^{2}$, the value of $d\sigma/dt$ remains almost unchanged
from ISR to LHC energies, i.e. at $t=t_{\ast \ast}\approx-2.3$ GeV$^{2}$ there is a second stationary point of $d\sigma/dt$ [54]. 
Unfortunately, we do not have experimental data for the energies between ISR and LHC in this $t$ range to see a more detailed picture of the energy 
dependence of $(d\sigma/dt)_{t=t_{2}}$, but we can expect behaviour similar to that of $(d\sigma/dt)_{t=t_{1}}$ in the forward diffraction cone.

The $t$-dependence of $d\sigma/dt$ in the region beyond the second maximum at fixed energy is well approximated by exponential form 
$C(s)\exp(-|t|D(s))$, where the slope $D(s)$ increases from 1.8 GeV$^{-2}$ at ISR energies up to 4.7 GeV$^{-2}$ at 13 TeV [3,7]. The factor $C(s)$
also grows with energy: from 2$\cdot$10$^{-3}$ mb/GeV$^{2}$ at ISR energies up to 1.7 mb/GeV$^{2}$ at 13 TeV. Just as for the forward diffraction 
cone (see Eq. 3), the energy dependence of $(d\sigma/dt)_{t=t_{2}}$ is given by 
\begin{equation}
 (\ln\frac{d\sigma}{dt})_{t=t_{2}}^{\prime} = (\frac{C^{\prime}}{C} - |t_{2}|D^{\prime}),
\end{equation}
where the prime denotes the derivative with respect to $ \ln (\sqrt{s})$. Due to an increase of $C(s)$, the ratio $C^{\prime}/C$ will decrease to zero
at high energies, therefore, $(d\sigma/dt)_{t=t_{2}}$ will decrease at sufficiently high energies for any fixed value of $|t_{2}|$ 
(if $D^{\prime}\geq const $ or decreases more slowly than $C^{\prime}/C$), as it does now for $|t|>2.3$ GeV$^{2}$.

Thus, the differential cross-section in the region beyond the second maximum reveals the same properties as in the forward diffraction cone:
structureless almost exponential decrease with $|t|$ at a fixed energy, growth of the factor $C(s)$ and slope $D(s)$ with energy,
very specific dynamics of energy dependence at a fixed $|t|$ (with a transitory stationary point at $t_{\ast \ast}$). In other words,
in the region beyond the second maximum, we observe the second diffraction cone of $d\sigma/dt$, which has the same origin as the forward
diffraction cone [54].

The simplest natural explanation for this experimental fact is that the amplitude of high energy elastic scattering is the sum of two similar functions
\begin{equation}
 T(s,t)=A_{1}(s,t)+A_{2}(s,t),
\end{equation}
which have the same fundamental status, that is, the functions $A_{1}(s,t)$ and $A_{2}(s,t)$ have the same structure and differ only in the values of 
the parameters contained in this structure. This difference in parameter values leads to $|A_{1}(s,t)|\gg|A_{2}(s,t)|$ in the region of the first
diffraction cone and to $|A_{1}(s,t)|\ll|A_{2}(s,t)|$ in the region of the second diffraction cone. Since the differential cross-section is given by
\begin{equation}
 \frac{1}{\pi}\frac{d\sigma}{dt}=|T|^{2}=|A_{1}|^{2}+|A_{2}|^{2}+2|A_{1}||A_{2}|\cos(\varphi_{1}-\varphi_{2}),
\end{equation}
where $\varphi_{1}(s,t)$ and $\varphi_{2}(s,t)$ are the phases of $A_{1}(s,t)$ and $A_{2}(s,t)$, so $d\sigma/dt\sim|A_{1}|^{2}$ and 
$d\sigma/dt\sim|A_{2}|^{2}$ in the region of the first and second diffraction cones, respectively. The interference of $A_{1}(s,t)$ and $A_{2}(s,t)$
is significant for $d\sigma/dt$ only in the $t$ region, where $|A_{1}(s,t)|\approx|A_{2}(s,t)|$, i.e. in the dip-bump (or shoulder for $p\bar{p}$)
region, because 
\begin{equation}
 \frac{1}{\pi}\frac{d\sigma}{dt}=(|A_{1}|-|A_{2}|)^{2}+2|A_{1}||A_{2}|(1+\cos(\varphi_{1}-\varphi_{2})).
\end{equation}
For this reason, $d\sigma/dt$ has only one dip-bump (or shoulder) structure, because $|A_{1}(s,t)|$ and $|A_{2}(s,t)|$ are smooth structureless
functions. The nature of the interference pattern (dip-bump or shoulder) is determined by the value of the relative phase $(\varphi_{1}-\varphi_{2})$. 

The shrinkage of the first and second diffraction cones with increasing energy in the ISR--LHC energy range looks like their clockwise rotation 
around their stationary points at $t_{\ast}$ and $t_{\ast \ast}$, respectively. As a result, the dip-bump structure (the region in which  
$|A_{1}(s,t)|\approx|A_{2}(s,t)|$) moves to $t=0$, becomes narrower, and the values of $d\sigma/dt$ at points of this structure grow 
with energy (see Fig. 1).
\begin{figure}[t]
\centering
\includegraphics[height=8.5cm]{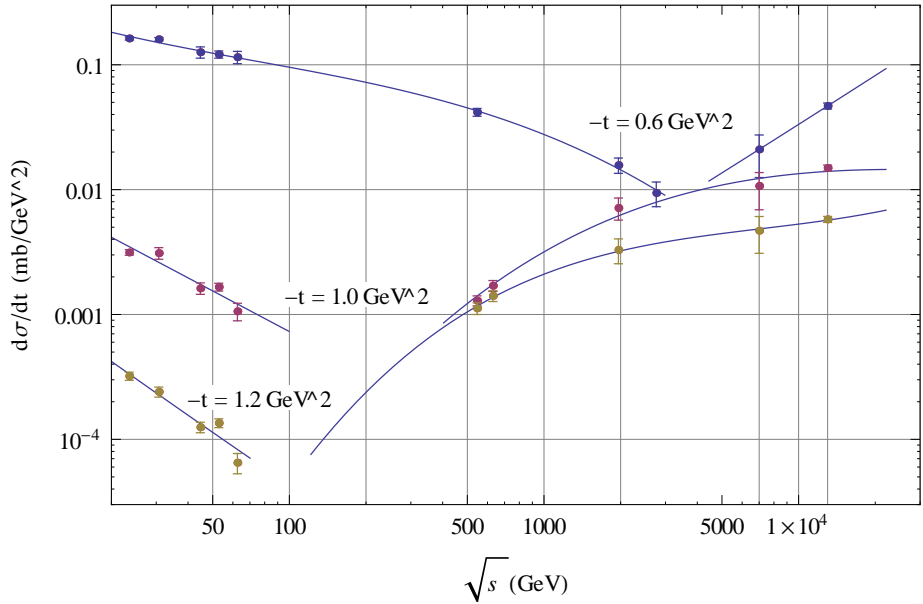}
\caption{Energy dependence of the elastic differential cross-section at fixed values of the momentum transfer $t$ in the interference region. 
Data are taken from Refs. [1-8] and [17-19,22]. The lines are the polynomial fits to the data points.}
\end{figure}

The energy dependence of $(d\sigma/dt)_{t=t_{3}}$ for fixed values of $t=t_{3}$ in the interference region reveals a transition from the decreasing 
part of the first diffraction cone to the increasing part of the second diffraction cone (see Fig. 7). This transition was first observed
at $t_{3}=-1.4$ GeV$^{2}$ in the ISR energy range: between 23 and 31 GeV $(d\sigma/dt)_{t=t_{3}}$ decreases, while between 45 and 62 GeV it is
already increasing [3]. At $t_{3}=-1.2$ GeV$^{2}$, the differential cross-section decreases (due to shrinkage of the first diffraction cone)
up to $\sqrt{s}\approx100$ GeV, where it has a local minimum (dip), and then increases due to the shrinkage of the second diffraction cone 
(see Fig. 7). At $t_{3}=-1$ and -0.6 GeV$^{2}$, the transition from the first to the second diffraction cone occurs at 200--300 GeV and $\sim3$ TeV, 
respectively. Thus, the experimental data in the interference region also confirm the above observations.

\section{The structural symmetry conjecture}

What are the underlying reasons for the non-trivial algebraic structure of the elastic scattering amplitude? Here we give one of the possible answers.

The elastic scattering amplitude depends not only on kinematic variables
\begin{equation}
x_{1}=s,\,\,\, x_{2}=t,\,\,\, x_{3}=u,\,\,\, x_{1}+x_{2}+x_{3}=\sum m_{i}^{2},
\end{equation}
but also on the set of parameters $a_{i}$ that characterize a given reaction channel (particle masses, quantum numbers, thresholds of open 
inelastic channels, etc.)
\begin{equation}
F_{1}=F(x_{1}, a_{1};\, x_{2}, a_{2};\, x_{3}, a_{3}).
\end{equation}
Due to the crossing symmetry, this function also describes the $t$- and $u$-channels
\begin{equation}
F_{2}=F(x_{2}, a_{1};\, x_{1}, a_{2};\, x_{3}, a_{3}),\,\,\, F_{3}=F(x_{3}, a_{1};\, x_{2}, a_{2};\, x_{1}, a_{3}).
\end{equation}
The permutation of $x_{2}$ and $x_{3}$ into $F_{1}$ gives the amplitude in the exchange channel 
\begin{equation}
F_{1ex}=F(x_{1}, a_{1};\, x_{3}, a_{2};\, x_{2}, a_{3}),
\end{equation}
and similarly for $F_{2}$ and $F_{3}$.

The three physical channels of the elastic scattering amplitude have the same status and differ only in the values of parameters, therefore, 
it is natural to assume that $F(z_{1},z_{2},z_{3})$ is a symmetric function under permutations 
$z_{i}=(x_{i}, a_{i})\longleftrightarrow z_{j}=(x_{j}, a_{j})$, $i,j=1,2,3$. If $F_{1}$ has such structural symmetry, then obviously $F_{2}$ and 
$F_{3}$ also have it. For example, the functions
\begin{equation}
(\frac{x_{1}}{a_{1}})^{n_{1}}(\frac{x_{2}}{a_{2}})^{n_{2}}(\frac{x_{3}}{a_{3}})^{n_{3}},\,\,\,
(\frac{x_{2}}{a_{1}})^{n_{1}}(\frac{x_{1}}{a_{2}})^{n_{2}}(\frac{x_{3}}{a_{3}})^{n_{3}},\,\,\,
(\frac{x_{3}}{a_{1}})^{n_{1}}(\frac{x_{2}}{a_{2}})^{n_{2}}(\frac{x_{1}}{a_{3}})^{n_{3}}
\end{equation}
are structurally symmetric functions. Another example of such a structurally symmetric function is
\begin{equation}
-G_{1}=\frac{\sqrt{1-a_{1}x_{1}}\sqrt{1-b_{3}x_{3}}}{(d_{13}+\sqrt{1-c_{13}x_{2}})^{n_{13}}}+
\frac{\sqrt{1-a_{3}x_{3}}\sqrt{1-b_{1}x_{1}}}{(d_{31}+\sqrt{1-c_{31}x_{2}})^{n_{31}}}+
\end{equation}
\begin{equation}
+\frac{\sqrt{1-a_{1}x_{1}}\sqrt{1-b_{2}x_{2}}}{(d_{12}+\sqrt{1-c_{12}x_{3}})^{n_{12}}}+
\frac{\sqrt{1-a_{2}x_{2}}\sqrt{1-b_{1}x_{1}}}{(d_{21}+\sqrt{1-c_{21}x_{3}})^{n_{21}}}+
\end{equation}
\begin{equation}
+\frac{\sqrt{1-a_{2}x_{2}}\sqrt{1-b_{3}x_{3}}}{(d_{23}+\sqrt{1-c_{23}x_{1}})^{n_{23}}}+
\frac{\sqrt{1-a_{3}x_{3}}\sqrt{1-b_{2}x_{2}}}{(d_{32}+\sqrt{1-c_{32}x_{1}})^{n_{32}}},
\end{equation}
where $a_{i},b_{i},c_{ij},d_{ij},n_{ij}$ are positive constants. The parameters in this example determine the location of the branch points and  
behaviour of $G_{1}$ for large values of $x_{1},x_{2},x_{3}$. Note that only the first two terms of $G_{1}$ are important at high energies 
and fixed values of $t$, i.e. in the diffraction region. The other four terms in this region are much smaller.

In the general case, the structurally symmetric amplitude is a composition of irreducible representations of the permutation group $S_{3}$:
\begin{equation}
F=T^{(123)}(R_{1}^{(23)}(E_{12}^{3}+E_{13}^{2})+R_{2}^{(13)}(E_{21}^{3}+E_{23}^{1})+R_{3}^{(12)}(E_{31}^{2}+E_{32}^{1})),
\end{equation}
where $(123)$ and $(ij)$ denote symmetry under permutations 1,2,3 and $i,j$ respectively. Functions in the above TREE-structure must have
correct analytic properties with respect to kinematic variables $x_{1},x_{2},x_{3}$, while the amplitude $F$ must satisfy all the constraints
which follow from analyticity and unitarity [63,64] for each of the three channels. These requirements restrict the arbitrariness in the choice of 
functions for the TREE-amplitude. For example, the simplest symmetric form $-\sqrt{4-x_{1}}\sqrt{4-x_{2}}\sqrt{4-x_{3}}$ is incompatible 
with unitarity. The elastic scattering amplitude should increase with $t$ in the region of about $t=0$ [23-27]. The simplest non-forbidden form is
$f_{1}g_{3}/h_{2}\equiv f(z_{1})g(z_{3})/h(z_{2})$, where $f,g,h$ are some growing functions of their kinematic variables and $z_{i}=(x_{i}, a_{i})$.
Then, in the corresponding TREE-form
\begin{equation}
\frac{f_{1}g_{3}}{h_{2}}+\frac{f_{3}g_{1}}{h_{2}}+\frac{f_{2}g_{3}}{h_{1}}+\frac{f_{3}g_{2}}{h_{1}}+\frac{f_{1}g_{2}}{h_{3}}+\frac{f_{2}g_{1}}{h_{3}},
\end{equation}
only the first two terms dominate at high energies in the diffraction region. It can be argued that any more or less realistic example of 
TREE-amplitude gives a two-component structure of the high energy amplitude in the diffraction region. We hope that this property of TREE-amplitude 
will be proved in the general case.

Next, we formulate the \textit{structural symmetry conjecture (SSC)}. At energies beyond the resonance region, the elastic scattering amplitude is
approximated by the TREE-form (16) with a finite number of parameters into which all three physical channels enter on the same footing. 
The transition to the $t-$ and $u-$ channels is set by the permutation $x_{1}\longleftrightarrow x_{2}$ and $x_{1}\longleftrightarrow x_{3}$
respectively. The TREE-amplitude has the correct analytic properties and satisfies all the constraints that follow from the general principles of the  
local QFT for each of the three channels. All six terms of the TREE-amplitude are essential at low and medium energies, but only two of them dominate 
at high energies in the diffraction region, meaning the other four terms are small in this region. At high energies and 
$|x_{1}|\sim|x_{2}|\sim|x_{3}|$, the TREE-amplitude has a power dependence in accordance with the dimensional counting rules. All terms of the 
TREE-amplitude are complex functions, that is, they have real and imaginary parts and, obviously, contain both C-even and C-odd parts.

This conjecture provides a natural explanation for the existence and similarity of two diffraction cones in $(d\sigma/dt)$ as a consequence 
of the ``democracy'' of physical channels in the amplitude structure. This explains why the terms $A_{1}(s,t)$ and $A_{2}(s,t)$ in Eq. (5) have the 
same diffraction status. Of course, to verify the above conjecture, it is necessary to construct a realistic TREE-amplitude, which will describe 
all the available $pp$ and $p\bar{p}$ data beyond the resonance region. This will be done in the future, we hope. Due to its general nature, 
the $SSC$ must be applied to all elastic processes, as well as to any reaction $2\longrightarrow2$.

\section{Summary and discussion}

We have analyzed the energy dependence of the differential cross-section for $pp$ elastic scattering in the ISR--LHC energy range at fixed values of
the momentum transfer. The available experimental data reveal a striking similarity between the behaviour of $(d\sigma/dt)$ in the region beyond
the second maximum and its behaviour in the forward diffraction cone. In particular, in this region $(d\sigma/dt)$ has a stationary point at 
$t=t_{\ast \ast}$, as for $t=t_{\ast}$ in the forward diffraction cone. This suggests that there are two diffraction cones of the same origin. The simplest way to explain these experimental facts is that the amplitude of high energy elastic scattering has a two-component structure with similar terms that differ only in the values of the parameters. The energy dependence of $(d\sigma/dt)$ at a fixed $t$ in the dip-bump region confirms that in this region 
there is a transition from the first shrinking diffraction cone to the second. We have presented arguments in favour of a non-trivial algebraic 
structure of the amplitude as a manifestation of the fact that the $s-,t-,u-$ channels enter the amplitude on an equal footing.  
As a result, we suggested a structural symmetry conjecture that explains the existence and similarity of two diffraction cones in $d\sigma/dt$ 
and provides a general recipe for constructing an adequate model for the elastic scattering amplitude.

Due to the unitarity condition, elastic scattering is largely a shadow process, that is, the properties of the elastic scattering amplitude 
are largely determined by inelastic interactions. In addition, unlike inelastic processes, elastic scattering has no probabilistic interpretation; 
we can only talk about the probability of survival of colliding hadrons (elastic scattering plus passage without interaction). The latter does not 
allow us to trust simple visual considerations, because our imagination has a semiclassical, probability-based character. For these reasons, 
elastic scattering is one of the most difficult problems in high energy physics, and we do not yet have an adequate model for this phenomenon. Indeed,
all important experimental discoveries in this field, such as the growth of $\sigma_{\mathrm{el}}(s)$, $\sigma_{\mathrm{tot}}(s)$, 
$\sigma_{\mathrm{el}}/\sigma_{\mathrm{tot}}$, the existence of the second diffraction cone, were unexpected, surprising for us.

We suggested the structural symmetry conjecture to explain the two-cone structure of $(d\sigma/dt)$ in the diffraction region, but $SSC$, as a
general idea, has many other consequences and applications. For example, according to the $SSC$, at high energies the total cross-section also has 
a two-component structure $\sigma_{\mathrm{tot}}(s)=\sigma_{1}(s)+\sigma_{2}(s)$. At low and medium energies, it is necessary to have a multicomponent
amplitude, as is the case, for example, in the Regge model. At high energies, it is necessary to have a two-component amplitude with similar
terms, say, the Pomeron-Odderon structure $A_{1}(s,t)$, plus another Pomeron-Odderon structure $A_{2}(s,t)$. In the impact parameter representation, the high energy amplitude also has two components,
but this manifestation of amplitude symmetry is not related to the structure of hadrons. The two-cone structure of $(d\sigma/dt)$ in the diffraction 
region is also observed for elastic scattering $\pi p$ and $Kp$ at high energies [65]. Thus, we can anticipate that the $SSC$ has a universal
character and can be applied to all elastic processes (and to any reaction $2\longrightarrow2$). For example, the $\pi p$ TREE-amplitude will describe
not only the elastic channels $\pi^{-}p\rightarrow\pi^{-}p$, $\pi^{+}p\rightarrow\pi^{+}p$, but also $p\bar{p}\rightarrow\pi^{+}\pi^{-}$.

\section{Acknowledgements}

I am grateful to V.A. Petrov, V.V.~Ezhela and N.P.~Tkachenko for useful discussions and the critical remarks.

\end{document}